\documentclass[fleqn,usenatbib]{mnras}
\usepackage{newtxtext,newtxmath}
\usepackage[T1]{fontenc}
\usepackage{ae,aecompl}

\usepackage{graphicx}	
\usepackage{amsmath}	

\newcommand\tsc{t_\text{sc}}         
\newcommand\dnusc{\Delta\nu_\text{sc}}      
\newcommand\tobs{t_\text{obs}}       

\newcommand\Fon{F_\text{on}}         
\newcommand\Foff{F_\text{off}}       
\newcommand\sigmaCoff{\sigma^\text{cs}_\text{off}}   
\newcommand\sigmaoff{\sigma_\text{off}}   
\newcommand\phiprime{\phi^{\,\prime}}
\newcommand\tbeg{t_\text{beg}}
\newcommand\tend{t_\text{end}}
\newcommand\TEC{C_\text{e}}
\newcommand\Ne{n_\text{e}}
\newcommand\nuplasm{\nu_\text{p}}
\newcommand\Deltanusp{\Delta\nu_\text{sp}} 
\newcommand\Aphi{A_\phi} 
\newcommand\Tphi{T_\phi} 
\newcommand\AD{A_{\text D}} 
\newcommand\Dm{D_{\text m}} 
\newcommand\mrmd{\text{d}}     

\title[Ionospheric effects in VLBI]{Ionospheric effects in
  VLBI measured with space-ground interferometer
  RadioAstron} 

\author[M. V. Popov et al.]{
M. V. Popov,$^{1}$
N. Bartel,$^{2}$
M. S. Burgin,$^{1}$\thanks{E-mail: mburgin@asc.rssi.ru}
T. V. Smirnova,$^{3}$
and V. A. Soglasnov$^{1}$
\\
$^{1}$Lebedev Physical Institute, Astro Space Center,
   Profsoyuznaya 84/32, Moscow, 117997, Russia\\
$^{2}$York University, 4700 Keele St., Toronto, ON M3J 1P3, Canada\\
$^{3}$Lebedev Physical Institute, Pushchino Radio
   Astronomy Observatory, Pushchino 142290, Moscow region, Russia
}

\date{Accepted 2021 July 01. Received 2021 June 23; in original form 2021 April 16}

\pubyear{2021}

\begin{document}
\label{firstpage}
\pagerange{\pageref{firstpage}--\pageref{lastpage}}
\maketitle

\begin{abstract}
We report on slow phase variations of the response of the
space-ground radio interferometer RadioAstron during
observations of pulsar B0329+54. 
The phase variations are
due to the ionosphere and clearly distinguishable from
effects of interstellar scintillation.
Observations were
made in a frequency range of 316-332~MHz with the 110-m
Green Bank Telescope and the 10-m RadioAstron telescope in
1-hour sessions on 2012 November 26, 27, 28, and 29 with
progressively increasing baseline projections of about 60,
90, 180, and 240 thousand kilometres.  Quasi-periodic phase
variations of interferometric scintles were detected in two
observing sessions with characteristic time-scales of 12 and
10 minutes and amplitudes of up to 6.9~radians.  We
attribute the variations to the influence of medium-scale
Travelling Ionospheric Disturbances.  The measured amplitude
corresponds to variations in vertical total electron content
in ionosphere of about $0.1\times10^{16}\, \text m^{-2}$. Such
variations would noticeably constrain the coherent
integration time in VLBI studies of compact radio sources at
low frequencies.
\end{abstract}


\section{Introduction}
When a source of intrinsically small angular size, such as a
pulsar, an AGN, or a component of an interstellar maser, is
observed at sufficiently low frequency, experimentally
measured position, the integrated brightness, size, and the
structure of its image are influenced and in many cases are
completely determined by the scattering by density
fluctuations in the intervening plasma.
Because of the motion of non-uniform plasma across the line
of sight and stochastic variability of the electron density
due to the turbulence, the scattered image exhibits both
temporal and angular random variations. 
Statistical properties of the time-varying image contain
information on the intrinsic structure of the observed
object and on the properties of the scattering plasma.

The only method capable of directly resolving the scattering
disk formed due to interaction of the radio emission with
the turbulent interstellar medium is 
very long baseline interferometry (VLBI). The results of
VLBI observations of interstellar scattering (ISS) may be
presented as the cross-spectrum
\begin{equation}  \label{dyncrosp}
\tilde{V}_{AB}(\nu ,t)=\tilde{E}_A(\nu,t){\tilde E}^*_B(\nu,t)
\end{equation}
of frequency $\nu$ and time $t$ produced by cross-correlating
data obtained on a pair of antennas $A$ and $B$.  Here
$\tilde{E}_A(\nu,t)$ is the Fourier transorm at frequency $\nu$
of the complex form of the electric field measured by
antenna $A$ and $\tilde{E}_B^*(\nu,t)$ is the corresponding
complex conjugate measured by antenna B.  In the special
case when $A=B$, the cross-spectrum reduces to the dynamic
spectrum, which is the real-valued function that describes
the observed flux density as a function of frequency and
time.  In the general case when $A\neq B$, the value of
$\tilde{V}_{AB}$ is complex, so to fully employ the
capabilities of VLBI, in comparing observations with
theoretical models it is necessary to take into account both
the amplitude, $|\tilde V|$, and the phase, $\arg \tilde V$,
of the cross-spectrum.

In many cases, however, the large systematic errors in
measured phase make it impossible to use it in the analysis.
One source of uncertainty is the variability of the
ionospheric excess path with the time scale comparable to
the duration of observations. To minimise its adverse effect
on the VLBI experiments aimed at investigating the
interstellar scattering, it is necessary to determine the
relevant parameters of the ionosphere at the time of
observations.

In the present paper, we report the measurements of the
phase variations in cross-spectra obtained from
observations of PSR B0329+54 with space-ground
interferometer RadioAstron.  Analysis of the visibility
amplitude measured during these observations have led
\citet{popov2016} and \citet{popov2017} to resolution of the
fine substructure in the scattering disk. It seems plausible
that the contribution of the substructure elements to the
observed cross-spectrum varies on the time-scale of the
order of the scintillation time, which leads to the
variability of the cross-spectrum phase.  The detection of
phase fluctuations caused by variability in the substructure
would yield additional information on the properties of the
scattering plasma.

The time dependence of complex
  cross-spectra -- both in amplitude and phase -- for several
  pulsars including PSR B0329+54 were used by
  \citet{2020ApJ...888...57P} as an initial point in their
  study of the intrapulse variations of interferometric
  visibility. But the variations with time-scale
  $\approx10^{-5}\,$s studied in that paper are $\approx10^6$ times
  faster than the ISS variability for the pulsar, and the
  origin of the variations is not related to
  non-stationarity of the intervening plasma.

In determining the cross-spectrum phase
variations due to ISS it is necessary to subtract the
ionospheric contribution from the observed values.  Ideally,
the ionospheric contribution should be calculated based on
parameters of the ionosphere measured synchronously with the
VLBI observations by an independent method, e.g., applying
the procedure described by \cite{Ros2000}, who used the GPS
data. 

In our case, independent measurements were not
available, and the only possible approach was to try to
derive the ionospheric contribution from our data alongside with
the phase variations due to interstellar
scattering.  The two effects can in principle be decoupled
by taking advantage of their different frequency dependence.
Thus, VLBI observations of pulsars potentially can yield
information both on ISS and on the phase distortions induced
by the ionosphere.  Since one of the antennas used in our
experiment was located beyond the atmosphere, the measured
ionospheric phase shift can be directly related to the state
of the ionosphere above the ground antenna.
\section{Observations} \label {sec:obs}
PSR B0329+54 is the brightest pulsar in the northern
hemisphere.  It is located at the outer edge of the Orion
spiral arm. Its Galactic coordinates are $l=145^\circ,\;
b=-1\fdg2$, and its parallax distance from the Sun is about
1~kpc \citep{brisken2002}.  We observed the pulsar in P-band
during four successive sessions on 2012 November 26, 27, 28,
and 29 with the 10-m RadioAstron space radio telescope (SRT)
and the 110-m Robert C. Byrd Green Bank Telescope (GBT). The
baseline projections were about 60, 90, 180, and 240
thousand kilometres for the four consecutive days,
respectively.

The RadioAstron mission and technical parameters of the SRT
have been described by \cite{kardashev2013}.  The SRT
recorded the data in the upper sideband of 316-332~MHz with
one-bit digitising and transmitted them in real-time to the
telemetry station in Puschino.  At the GBT the science data
were recorded with the MkVb VLBI recording system with
two-bit digitising. In order to reduce the influence of
strong bandpass distortion at GBT near the high-frequency
edge of the sideband, we used only the measurements in the
range 316-330~MHz.

Each observing session lasted one hour. The automatic level
control, pcal, and noise diode were turned off to avoid the
distortion of the pulsar flux measurements.  We recorded the
data in 570 second scans with 30 second gaps between them,
but only used data obtained during the first 500 seconds of
each scan in the analysis.  The recorded data were
transferred via internet to the Astro Space Center (ASC) in
Moscow.

\section{Data reduction} \label{sec:red}

For quantitative analysis, we selected the observations
performed on November 26 and 29.  The choice stemmed from
two considerations.  First, we tried to maximize the spread
of baseline projections. Second, the exploratory inspection
of data revealed that phase variations are
significantly stronger on November 26 and 29 than on two
other dates.

The data were processed in two steps.  At the first
stage, which relied mainly on GBT data, we computed the
average pulse profile and principal characteristics,
the scintillation time and decorrelation bandwidth, of the
interstellar scintillations (ISS).  The results derived at this
stage were used to set some parameters of algorithms applied
at the next step of processing.

The second stage comprised evaluation of the cross-spectrum
for the baseline SRT-GBT and investigation of the temporal
variations of the cross-spectrum phase, which is directly
influenced by the ISS and ionosphere.

\subsection{Pulse profile and dynamic spectra} \label{sec:DynSp}

\begin{figure*} 
\includegraphics{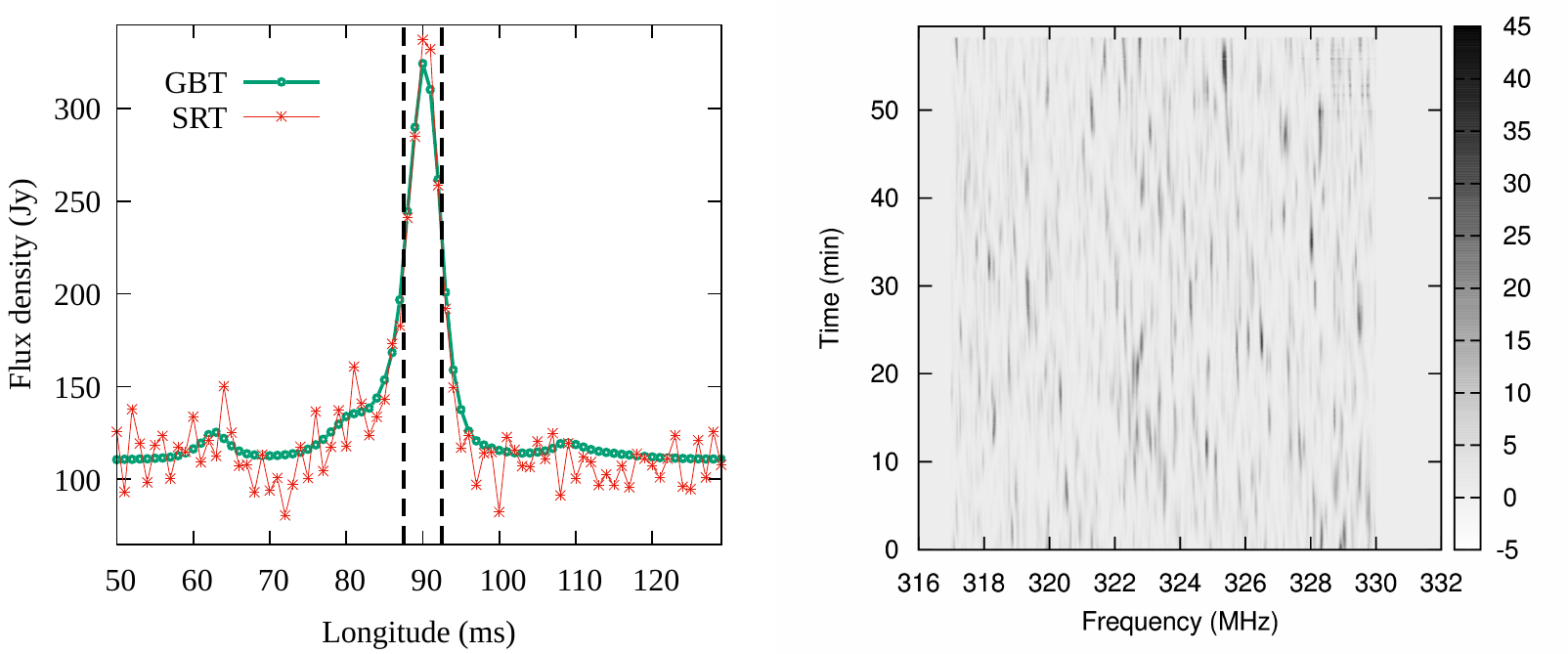}    
\caption{Left-hand panel: average profile obtained in an
  observing scan. The vertical dashed lines delineate
  the window used for calculation of the on-pulse dynamic
  spectra and cross-spectra.  Right-hand panel:
  dynamic spectrum based on GBT observations for the
  session on November~26.
  \label{fig:prof_and_DSP} 
}
\end{figure*}

The first step of the analysis was to form the average pulse
profile. The pulsar period of 0.714 s was divided in
longitude (pulsar rotational phase) bins of 1~ms
duration. Each observed instantaneous value of flux density
was assigned to a bin according to the longitude calculated
using TEMPO2 \citep{Hobbs2006}. Then, for each bin, the
corresponding measurements were averaged over a scan.  The
left-hand panel of Fig.~\ref{fig:prof_and_DSP} shows a
portion of the average profile obtained in an observing scan
at the GBT and the SRT.  The sensitivity and the noise level
of the SRT were calibrated using the GBT measurements as the
reference.

The data were correlated with the ASC correlator
\citep{likhachev2017} using a frequency resolution of
$\approx3.9$\,kHz with dedispersion and gating.  The produced
dynamic and cross-spectra were sampled synchronously with the
pulsar period. In this paper, we used the results of
correlation for two longitude windows each of 
5~ms duration: the on-pulse window, shown in
Fig.~\ref{fig:prof_and_DSP}, and the off-pulse window
located at the half-period offset from the main pulse, that
is beyond the longitude range of the figure.
The results of the correlation were stored in standard
FITS-IDI format \citep{2016FITS-IDI}.  The CFITSIO package
\citep{Pence1999} was used for the subsequent analysis.

In the next stage of data processing, we applied corrections
for the receiver bandpasses, performed the clean-up of the
spectra to remove strong narrow-band interferences, and 
normalized the pulse intensities in order to reduce the
influence of strong pulse-to-pulse intensity fluctuations. 
The normalization technique was described by
\citet{popov2017}.

The right-hand panel of Fig.~\ref{fig:prof_and_DSP} shows an example
of the resulting dynamic spectrum $S(\nu,t)$ calculated from the
data obtained at GBT on November~26. To construct the dynamic
spectrum for the whole one-hour session we filled gaps
between the scans by interpolating between the last and the
first spectra of consecutive scans.

The interstellar scintillation causes strong variations of
the observed flux density and the dynamic spectrum may be regarded
as an ensemble of bright spots on the time-frequency plane
called ``scintles''.  The primary statistical characteristics
of the ensemble are the decorrelation bandwidth, $\dnusc$,
and the scintillation time, $\tsc$, that describe the
average extent of the scintles in the respective directions.

For the observations used in the present paper, $\dnusc$ and
$\tsc$ were evaluated by \citet{popov2016}.  From the
analysis of the two-dimensional cross-correlation between
the dynamic spectra for the LCP and RCP (left-hand and
right-hand circular polarization) channels of GBT they
obtained values of $\dnusc\approx15$~kHz and $\tsc\approx114$~s.

\subsection{Phase variations in cross-spectra} \label{sec:PhaseVar} 

Our further analysis is based on the cross-spectra produced
at the previous stage of data processing.  To increase the
signal-to-noise ratio, we averaged the on-pulse
cross-spectra over ten consecutive pulsar periods, that is
over $\tobs\approx7.14$~s.  In the course of averaging, we took
into account only pulses that satisfy the condition
$\Fon\ge\left<\Foff\right>+6\sigmaoff$.  Here $\Fon$ is the
flux density in the on-pulse window,
$\left<\Foff\right>$ and $\sigmaoff$ are the mean and
standard deviation of the off-pulse flux density.  In rare
cases where none of the pulses falling into the range of
averaging meets the condition we assume that the
cross-spectrum is equal to the previous one.

\begin{figure}
\includegraphics{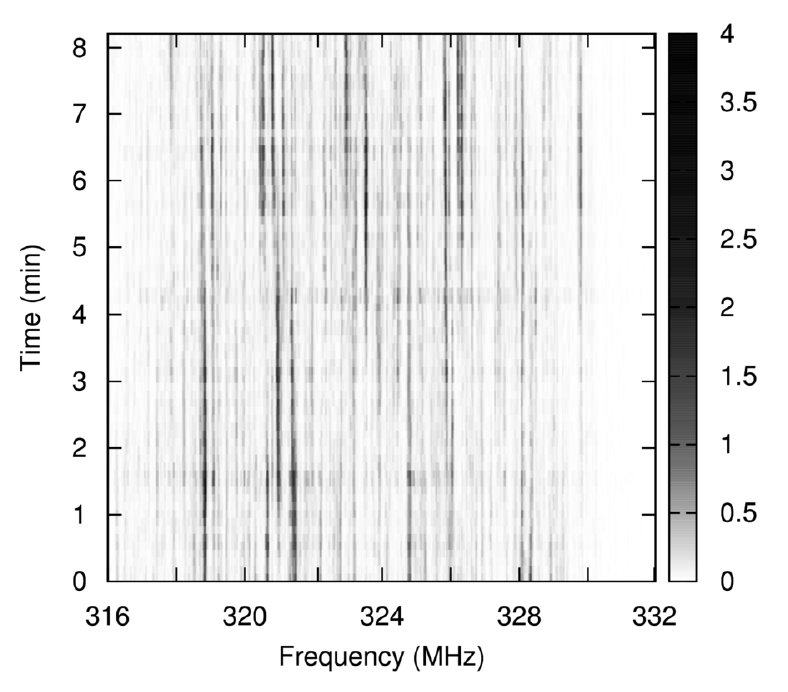}    
\caption{ Time-averaged correlated flux density on the
  baseline SRT-GBT during a scan on November 26 in RCP
  channel.  The averaging interval is $\approx7.14$~s, correlated
  flux density is in correlator units.
  \label{fig:frrt_freq}
} 
\end{figure}

Fig.~\ref{fig:frrt_freq} shows an example of the averaged
cross-spectrum. Its general structure originating from the
ISS closely resembles the structure in the dynamic spectrum
discussed in the previous subsection.  Since the averaging
time, $\tobs$, obeys the condition $\tobs\ll\tsc$, each
scintle contains, on average, $\tsc/\tobs\gg1$ visibility
measurements at different points in time, which make it
possible to study the temporal variability of the complex
cross-spectrum phase within individual scintles.  As
discussed in Section~\ref{sec:Origin} below, the comparison
between phase variations within different scintles permits
to disentangle the ionospheric contribution to the phase
shift from the effects of ISS.

The scintles used in our analysis were constructed as
follows. To increase the SNR we averaged the complex
cross-spectra produced by the correlator over eight
channels, thus reducing the frequency resolution to
31.25~kHz (about $2\times \dnusc$). 

\begin{figure*}
\includegraphics{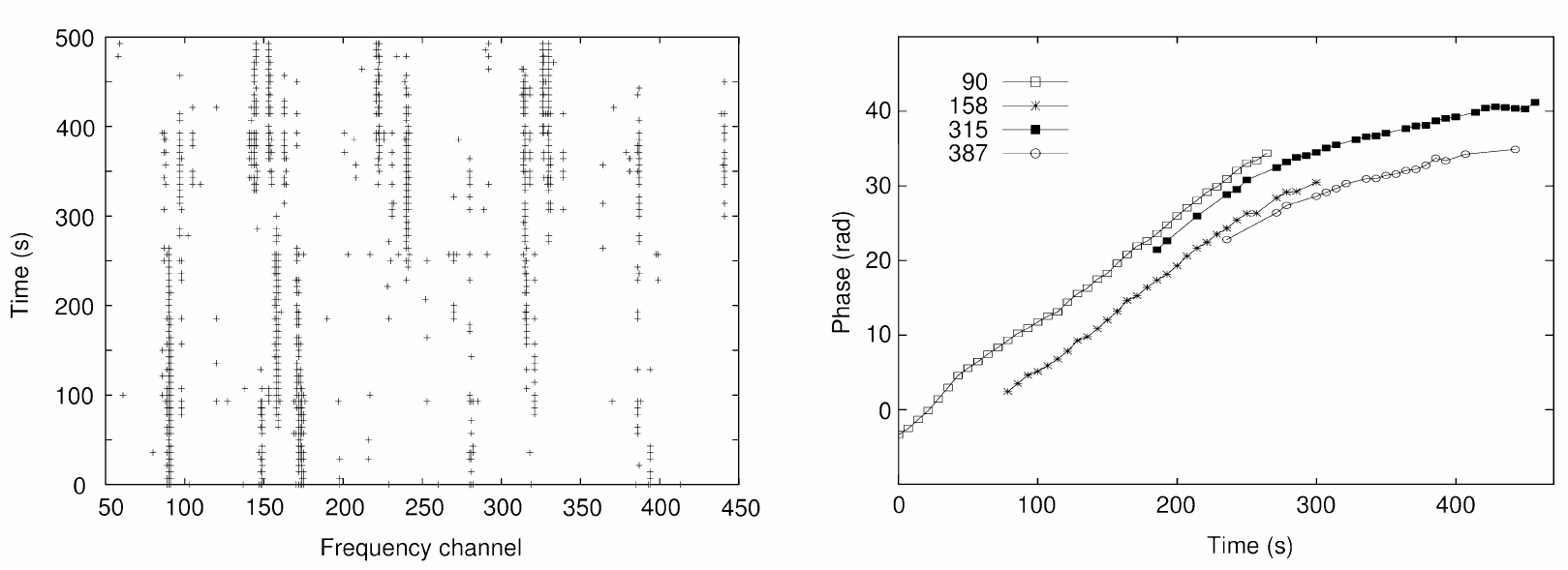}    
\caption{Left-hand panel: schematic presentation of the
  pixels selected for tracing the cross-spectrum phase
  variability for the scan started on 2012 November 26 at
  21:00 UTC.  The plus symbols represent bright pixels that
  constitute the selected scintles.  Right-hand panel: phase
  variations for the selected scintles.  The numbers in the
  legend specify the numbers of corresponding frequency
  channels in the left panel.  
  The nearly parallel signature of
  the changes of phase for all scintles is interpreted as
  being due to the ionosphere. The offsets between the
  phases of the scintles may be due to characteristics of
  ISS.
\label{fig:scintle}
}
\end{figure*}

  The pixels of the averaged
  cross-spectra that exceed in visibility magnitude the
  threshold of $3.5\sigmaCoff$ are considered ``bright'' and
  form a constituent part of a scintle.  Here, $\sigmaCoff$
  is the rms amplitude of the cross-spectrum measured in the
  off-pulse window.  The pixels with the visibility
  magnitude below the threshold are considered ''dim'' and
  the values of their phases are not used in further
  analysis.
  The left-hand panel of Fig.~\ref{fig:scintle} illustrates
  the results of the application of the algorithm described
  above to the scan started on 2012 November 26 at 21:00
  UTC.

  The pixels are then grouped into scintles in such a way
  that 
  (a) all pixels
  of a scintle reside in the same averaged frequency
  channel and cover a contiguous time interval, 
  (b) first and last pixels of a scintle, as well as 
  the majority, but not necessarily all, of intermediate
  pixels are bright, and
  (c) the time interval between two adjacent bright pixels
  located in the same scintle does not exceed $\tsc/2$.

The right-hand panel of
  Fig.~\ref{fig:scintle} illustrates the phase as a function
  of time for the four strongest scintles. The initial phase
  of $i$-th scintle is defined only within an ambiguity of
  $2n_i\pi$, where $n_i$ is an arbitrary integer, and
  therefore only comparison of the slopes of the curves in
  the plot is meaningful.  In creating the plot, the values
  $n_i$ were chosen in such a way as to facilitate such a
  comparison.  The apparent offsets between the phase curves
  for the different scintles could be due to characteristics
  of the ISM but a more detailed analysis is not possible
  because of the $2\pi$ ambiguities and the relatively short
  observation times.

The phase
variations for the four scintles, and in fact for all other
inspected scintles, are well approximated by a piecewise
linear function with the slope brake at $\approx300$~s.  For all
the scintles the rate of change of the phase is nearly the
same during the whole duration of the scan.  The same
analysis was performed for other  scans and in all cases phase
differences $\phi_i(t)-\phi_j(t)$ for
any pair $i,\,j$ of scintles indices
were constant over time
within an accuracy of our measurements.

The disadvantage of this method is that  removing
$2\pi$ ambiguities becomes non-unique for long
intervals between consecutive bright pixels.  Because of
the inter-scan gaps, this difficulty inevitably arises
in trying to study the variability on the time-scales
exceeding the duration of the scan.

To investigate the phase behaviour over longer times, we
used a different approach that relies only on the instant
phase shift rate, $\phiprime=\mrmd \phi/\mrmd t$, which we
approximate by the phase difference between bright pixels
separated in time by one $\approx7.14$~s long averaging
interval. This approach permits us to analyze the data
without having to correct for the $2\pi$ phase ambiguities.

\begin{figure*}   
\includegraphics{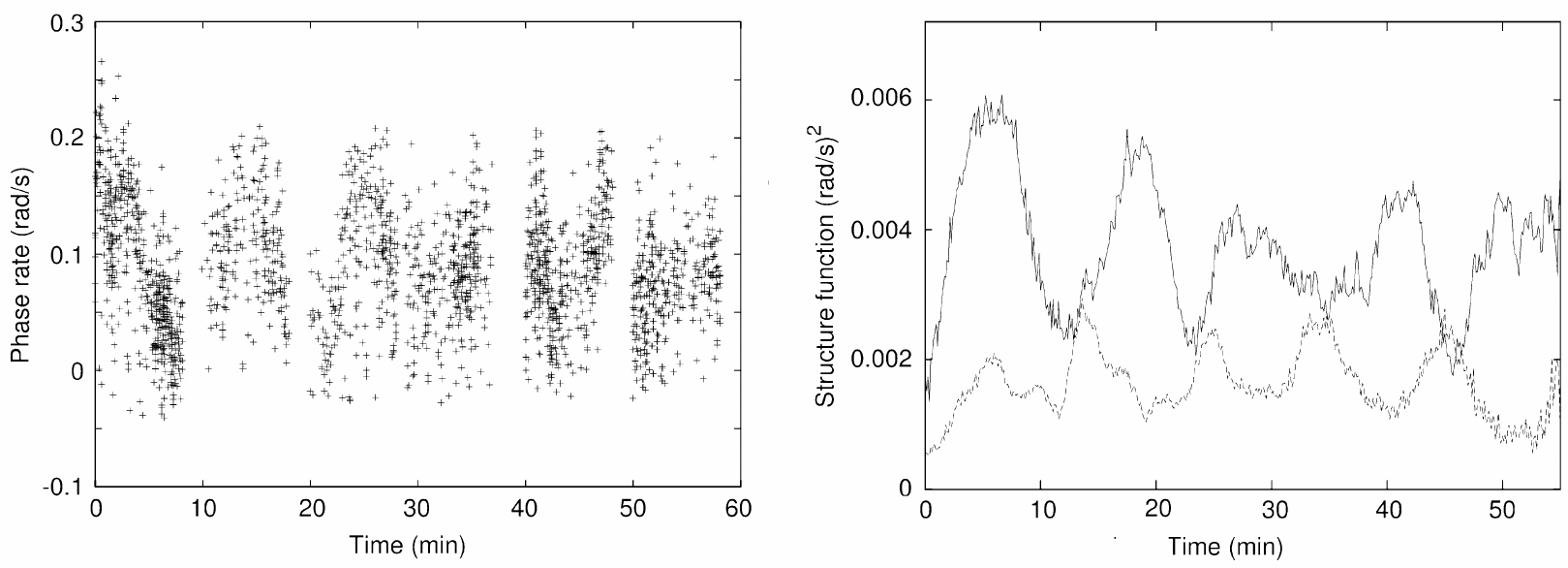}    
\caption{Left-hand panel: phase shift rate variation with time for
  all scintles in the one-hour session on November 26.
  Right-hand panel: structure functions for the phase shift rate on
  November 26 (solid line) and on November 29 (dashed line).
  \label{fig:phase_rate}
} 
\end{figure*}

The left-hand panel of
Fig.~\ref{fig:phase_rate} shows the phase shift rate as a function
of time during the session on 2012 November 26.  
As is evident from the left-hand panel of the figure, the
average value of phase shift rate, $\langle\phiprime\rangle$, 
over, say, 60 min is clearly larger than zero, meaning that 
$\phi$ gradially increases during the observations.

Another feature easily seen on the plot is a quasi-periodic
oscillation of $\phiprime$ around its average value, $\langle\phiprime\rangle$.
Our analysis of the oscillations is based on
the temporal structure function of phase shift rate, $D(\tau)$, defined as
\begin{equation}       \label{eq:defstrfun}
D(\tau)=\left\langle\left[\phiprime(t+\tau)-\phiprime(t)\right]^2\right\rangle\,.
\end{equation}
Here $\left\langle\right\rangle$ denotes averaging over the time interval
$[\tbeg,\tend-\tau]$ and over all selected scintles, where $\tbeg$ and
$\tend$ are the moments of the beginning and end of the session,
respectively. If we assume that slow gradual changes in $\phi$
can be approximated by a linear dependence on $t$, then
$D(\tau)$ is completely determined by the oscillatory component
of $\phiprime(t)$.

To establish the relationship between the amplitudes of
oscillation of $\phiprime(t)$ and $D(\tau)$, which we denote as
$\Aphi$ and $\AD$, respectively,  we consider the
simple model  where
$\phiprime_m(t)=\Aphi\cos(2\pi t/T)+\epsilon(t)$, and where
it is assumed that measurement errors $\epsilon(t)$ are normally
distributed random values 
with variance
independent of $t$, and
$\left<\epsilon(t)\epsilon(t^\prime)\right>=0$ for $t\neq t^\prime$.  Straightforward
calculation of the structure
function for the model, $\Dm(\tau)$, yields
$\Dm(\tau)=\AD(1-\cos(2\pi\tau/T))+\langle\epsilon^2\rangle$ where $\AD=\Aphi^2$.

The right-hand panel of Fig.~\ref{fig:phase_rate}
shows the structure functions for both observing sessions.
Approximating the data by the simple model described above
we may estimate the characteristic
period, $\Tphi$, and the amplitude, $\Aphi$, of the observed phase
shift rate variations
to be:
 $\Tphi\approx12$~min, $\Aphi\approx0.030$~rad/s
for the November 26 observing session, and $\Tphi\approx10$~min,
$\Aphi\approx0.015$~rad/s for the November 29 observing session.

\section{Origin of the observed phase variations}  \label{sec:Origin}

  Two principal effects that may
  cause temporal variability in cross-spectrum phase --
  interstellar scintillations and changes in parameters of
  the ionosphere -- are distinguishable from one another
  through their dependence on observation frequency.

In space-ground interferometry, the principal parameter of
the ionosphere that determines its influence on the observed
visibility phase is the total electron content (TEC) at the
location of the ground telescope.  TEC is defined as
\begin{equation}                                       \label{eq:TEC}
\TEC=\int_0^\infty\Ne(h)\,dh\,,
\end{equation}
where $\Ne(h)$ is the electron density at height $h$.
Since our observing frequency, $\nu$, satisfies the condition
$\nu \gg \nuplasm$, where $\nuplasm \loa 12$~MHz is the typical
plasma frequency of the ionosphere, we can use the
high-frequency approximation for the plasma refractive
index.  The ionosperically induced phase shift, $\Delta\phi$, is
then given by
\begin{equation}                           \label{Delta_phi}
 \Delta\phi=-8.45\frac{\TEC}{10^{16}\,{\text m}^{-2}}
          \frac{1\,\text{GHz}}{\nu} \sec z\,, 
\end{equation}
where $z$ is the zenith angle of the observed object 
\citep{Mevius2016,2017isra.book.....T}.

The scintles used in Section~\ref{sec:PhaseVar} for the
analysis of phase variability cover the frequency range,
$\Deltanusp$, of about 12~MHz (see
Fig.~\ref{fig:frrt_freq}).  
Ionospheric phase variations in
  all the scintles are governed by changes in the single
  parameter, $\TEC$. Expected relative difference in the magnitude
  of the slopes of dependence $\phi(t)$ for any two scintles
  is less than $\Deltanusp/\nu\approx0.04$ for
  the whole observing band and less then 0.03 for the
  brightest scintles presented in right-hand panel of
  Fig. \ref{fig:scintle}.
Consequently, the values
$\phi_i(t)-\phi_j(t)$, where $\phi_i(t)$ is the ionospheric phase
shift measured at the $i$-th scintle, remain nearly constant in
the course of a scan.

In contrast, phase variations due to ISS in widely
($\Delta\nu>\dnusc$) separated scintles are expected to be weakly
correlated, since different scintles originate from the
interference of the rays passing through different regions
of the scattering interstellar plasma.

  If we suppose that each cross-spectrum scintle originates
  from an element of the scattering disk substructure, then
  ISS contribution to temporal variations of observed phase
  differences $\phi_i(t)-\phi_j(t)$ reflects relative motion
  of the corresponding substructure elements.

  As it is illustrated in Fig.~\ref{fig:scintle}, our
  observations do not show any noticeable deviations from
  the synchronous phase changes in different scintles, which
  is compatible with the assumption that the observed
  temporal variations are dominated by the ionospheric
  contribution, and ISS effects are below the sensitivity of
  our method.

To explain the slowly varying component of the observed
temporal variability pattern, we note that the observations
were made in the time range 21:00 to 22:00 UTC, during the
last hour before the sunset at GBT. The telescope tracked
the source in the azimuth range {35--40\degr} and in
the range of elevation angles 20--27\degr.  The nearly linear
trend in phase can be attributed to diurnal variations in
TEC in the ionosphere above GBT and the decrease in $\sec z$ of
the ascending source.

Before proceeding to the interpretation of the
quasi-periodic phase variations, we estimate the amplitude
of changes in TEC above the GBT that correspond to the
amplitude, $\Aphi$, and time-scale, $\Tphi$, of the observed
oscillations of $\phiprime$.  With the amplitude of phase
variations, $\Delta\phi$, approximated by $\Delta\phi \approx \Aphi\Tphi/\pi$ and
the values for $\Aphi$ and $\Tphi$ from Section 3.2,
the amplitudes of TEC oscillations are $\approx0.1\times10^{16}$ and
$\approx0.05\times10^{16}\,\text{m}^{-2}$ on November 26 and 29,
respectively.

The observed amplitudes of TEC oscillations, time-scales of
quasi-periodicity, and low amplitude attenuation rates
indicate that the oscillations originate probably from
medium-scale travelling ionospheric disturbances (TID).
Ionospheric disturbances were studied in many publications
by various methods. Applications of VLBI to the problem
are described, e.g., by \cite{Zhi1993}, \cite{Hobiger2006},
\cite{Heinkel2009}, \cite{Helm2014}).  A review of important
aspects related to probing the ionosphere with RadioAstron
was given by \cite{Zhuravlev2020}.  It is generally accepted
that TIDs are a manifestation of atmospheric gravity waves
(AGW) in the ionosphere \citep{Hocke1996}.  A short review
of TIDs is provided by \cite{CROWLEY2018}. According to
them, there are several geophysical phenomena -- ocean waves,
tsunamis, explosions, weather fronts, and thunderstorms --
that might excite medium-scale AGWs, and consequently TIDs.

The typical time-scale of oscillations caused by
medium-scale TIDs is 10--30 minutes.  \cite{Azeem2017}
reported the observations of TIDs excited by earthquakes and
subsequent tsunami‐launched AGWs using GPS total electron
content (TEC) data. They found a maximum amplitude of
variations in TEC of $\approx1.1\times10^{16}\,{\text m}^{-2}$, and
a decay time of about four hours.  Thus, the phase
oscillations that we observed closely resemble the
variations induced by medium-scale TIDs.

\section{Conclusion}    \label{sec:Concl}

We present an analysis of two-element interferometry data in
the frequency range 316-332 MHz with the Earth–space
baseline GBT-SRT for PSR B0329+54.
  Analyzing interferometer phase
  variations in scintles produced by the interstellar
  medium, we show that ionospheric effects on the phase can
  be clearly identified and quantified.

  Cross-spectrum phases were monitored over time for several
  scintles. Variations that occurred synchronously in the
  scintles over the whole frequency range of the bandpass
  are attributed to variations of the ionosphere. Phase
  changes due to the effects of interstellar scintillation
  are expected to be uncorrelated for scintles widely
  separated in frequency.

The time dependence of the phase was calculated for
individual scintles.  Intercomparison of cross-spectrum
phases measured for multiple scintles permits to identify
the origin of phase variations. The variations arising from
the ionosphere occur synchronously in the scintles over the
whole frequency range, while phase changes due to effects of
interstellar scattering are expected to be uncorrelated for
the scintles widely separated in frequency.

We found that the observed phase variations are nearly
frequency independent, which means that they are dominated
by the ionospheric contribution.  Phase dependence on time
may be expressed as a sum of a slowly varying quasi-linear
component and quasi-periodic oscillations.  The slowly
varying component is naturally explained by the diurnal
variations of electronic density in the ionosphere and the
changes in the zenith distance of the pulsar as seen by the GBT. 

The oscillatory component corresponds to quasi-periodic
variations in the TEC in the
ionosphere above the GBT with amplitudes of (0.05--0.1)$\times
10^{16}\,\text{m}^{-3}$, time-scales of 10--12~min, and decay
times exceeding one hour. The TEC varions of this type are likely
caused by the medium-scale travelling ionospheric
disturbances.

  Interstellar scattering limits the coherent integration
  time, $\tobs$, for pulsar VLBI studies by the value of
  scintillation time, $\tsc$, which is typically from
  several seconds to several minutes \citep{goodman1989,
    gwinn1998}. These constraints do not apply to VLBI
  studies of extragalactic radio sources because their
  intrinsic angular sizes are large enough to quench
  diffraction scintillations in frequency and time
  \citep{1998MNRAS.294..307W}.  Therefore, only ionospheric
  and atmospheric effects would limit the coherent
  integration time for these sources.  In particular, the
  observed medium-scale Travelling Ionospheric Disturbances
  would pose the constraint $\tobs\ll\Tphi$ on the duration of
  low-frequency observations.

\section*{Acknowledgments} 
We thank the anonymous reviewer for helpful comments that
significantly improved the presentation of this work.
The RadioAstron project is led by the Astro
Space Center of the Lebedev Physical Institute of the
Russian Academy of Sciences and the Lavochkin Scientific and
Production Association under a contract with the Russian
Federal Space Agency, in collaboration with partner
organizations in Russia and other countries. 
Green Bank Observatory is supported by the National Science
Foundation and is operated by Associated Universities, Inc.

\section*{Data availability} 
The data underlying this article are available in
the RadioAstron data archive at
\url{ftp://ftp.radioastron.ru/raes10/},
and can be accessed with observation codes 
raes10a and raes10d

\bibliographystyle{mnras}
\bibliography{popov.r2}
\bsp	
\label{lastpage}
\end{document}